\documentclass[journal]{IEEEtran}
\usepackage{amsmath,amsfonts}
\usepackage{algorithmic}
\usepackage{algorithm}
\usepackage{array}
\usepackage[caption=false,font=normalsize,labelfont=sf,textfont=sf]{subfig}
\usepackage{textcomp}
\usepackage{stfloats}
\usepackage{url}
\usepackage{verbatim}
\usepackage{graphicx}
\usepackage{cite}
\usepackage{verbatim}
\usepackage{multirow}
\usepackage{makecell}
\usepackage{booktabs}
\usepackage{threeparttable}
\usepackage{amssymb}
\usepackage{bbding}
\usepackage{xcolor}

\begin{document}

\author{Ming~Cheng,
        Ming~Li, \IEEEmembership{Senior Member, IEEE}
\IEEEcompsocitemizethanks{
	\IEEEcompsocthanksitem Ming~Cheng and Ming~Li are with the School of Computer Science, Wuhan University, Wuhan 430072, China, and also with Suzhou Municipal Key Laboratory of Multimodal Intelligent Systems, Digital Innovation Research Center, Duke Kunshan University, Kunshan 215316, China.}
\thanks{Corresponding author: Ming Li, E-mail: ming.li369@dukekunshan.edu.cn}
}

\title{Multi-Input Multi-Output Target-Speaker Voice Activity Detection For Unified, Flexible, and Robust Audio-Visual Speaker Diarization}

\maketitle

\begin{abstract}
Audio-visual learning has demonstrated promising results in many classical speech tasks (e.g., speech separation, automatic speech recognition, wake-word spotting). We believe that introducing visual modality will also benefit speaker diarization. To date, Target-Speaker Voice Activity Detection (TS-VAD) plays an important role in highly accurate speaker diarization. However, previous TS-VAD models take audio features and utilize the speaker's acoustic footprint to distinguish his or her personal speech activities, which is easily affected by overlapped speech in multi-speaker scenarios. Although visual information naturally tolerates overlapped speech, it suffers from spatial occlusion, low resolution, etc. The potential modality-missing problem blocks TS-VAD towards an audio-visual approach. This paper proposes a novel Multi-Input Multi-Output Target-Speaker Voice Activity Detection (MIMO-TSVAD) framework for speaker diarization. The proposed method can take audio-visual input and leverage the speaker's acoustic footprint or lip track to flexibly conduct audio-based, video-based, and audio-visual speaker diarization in a unified sequence-to-sequence framework. Experimental results show that the MIMO-TSVAD framework demonstrates state-of-the-art performance on the VoxConverse, DIHARD-III, and MISP 2022 datasets under corresponding evaluation metrics, obtaining the Diarization Error Rates (DERs) of 4.18\%, 10.10\%, and 8.15\%, respectively. In addition, it can perform robustly in heavy lip-missing scenarios.
\end{abstract}


\section{Introduction}
\label{sec:intro}

\IEEEPARstart{L}{ike} documenting events in a diary, speaker diarization is the task of automatically detecting multiple speakers' utterance boundaries in conversational data~\cite{tranter2006overview}. It aims to split the audio or multi-modal signals into segments with labeled identities, solving the problem of ``Who-Spoke-When.'' As a front-end technique, it is essential in various downstream applications (e.g., speech recognition)~\cite{park2022review}.

In previous studies, speaker diarization research mainly focuses on audio streams~\cite{park2022review}. The conventional method, also known as the modularized method, utilizes cascaded modules to partition the audio signal into short segments and cluster their identities by advanced speaker representation techniques~\cite{wang2018speaker,park2019auto,landini2022bayesian}. These methods cannot handle overlapped speech as each audio segment is supposed to be speaker-homogeneous. Some studies propose additional post-processing techniques (e.g., overlapped speech detection~\cite{landini2020but}, overlap-aware resegmentation~\cite{bredin2021end,landini2022bayesian}) to compromise this effect. Then, End-to-End Neural Diarization (EEND) systems~\cite{fujita2019end_1,fujita2019end_2,horiguchi2020end,horiguchi2022encoder} are proposed to estimate multiple speakers' speech activities as multi-label classification. The end-to-end structure of neural networks leads to ease of optimization and robustness to overlapped speech. Nevertheless, permutation-invariant training in EEND-based methods causes performance degradation when the number of speakers increases in long audios. Although a few studies~\cite{kinoshita2021integrating,kinoshita2021advances,kinoshita2022tight,horiguchi2021towards,horiguchi2022online} have explored the unsupervised clustering to address this problem, their results are still unsatisfactory. Recently, TS-VAD approaches~\cite{medennikov2020target,wang2022similarity,cheng2023target,wang2023target} become attractive, which combine advantages of modularized methods and end-to-end neural networks. As a post-processing method, the TS-VAD framework requires an initial diarization system (e.g., modularized method) to extract each speaker's acoustic footprint as the speaker profile, solving the problem of ``Who-Spoke'' in advance. Then, a neural network-based model takes speech features and all speaker profiles to predict their corresponding framewise voice activities, aiming to address the ``When'' problem. This two-stage process has demonstrated excellent performance in popular benchmarks such as DIHARD-III~\cite{wang2021ustc} and VoxSRC21-23~\cite{wang2021dku,wang2022dku,cheng2023dku}.

However, speaker diarization in complex environments (e.g., far-field and highly overlapped speech, a large number of speakers) is still challenging. Using visual information as the complementary modality to improve diarization systems becomes a promising direction. Existing works mainly depend on constructing cross-modal synergy~\cite{el2014audiovisual,kapsouras2017multimodal,gebru2017audio,chung2019said}, clustering on audio-visual pairs~\cite{xu2022ava,wuerkaixi2022dyvise}, or end-to-end audio-visual diarization~\cite{he2022end,zhang2023flyspeech}, which are basically derived from the previous audio-only methods. Motivated by the highly accurate performance in TS-VAD studies, a question arises if it is feasible to investigate this framework in an audio-visual manner. Although similar works about Active Speaker Detection (ASD)~\cite{roth2020ava,tao2021someone,jiang2023target} also estimate the speaker's voice activities using audio-visual signals, they usually work for a single speaker at once and neglect the modality-missing problems. So far, there is a lack of an audio-visual diarization framework that can effectively deal with multi-speaker scenarios where the visual modality often suffers occlusion, off-screen speakers, or unreliable detection.

\begin{figure*}[t]
\centering
	\subfloat[Encoder-Only~\cite{medennikov2020target,wang2022similarity}]{\includegraphics[width=2.2in]{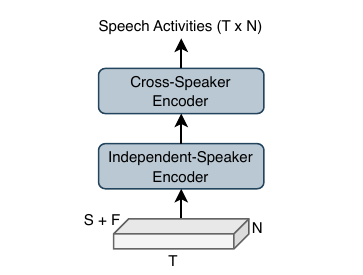}
	\label{fig_intro1}}
	\hfil
	\subfloat[Sequence-to-Sequence~\cite{cheng2023target}]{\includegraphics[width=2.2in]{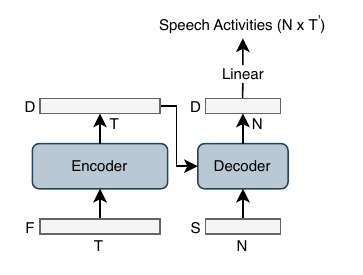}
	\label{fig_intro2}}
	\hfil
	\subfloat[Multi-Input Multi-Output]{\includegraphics[width=2.2in]{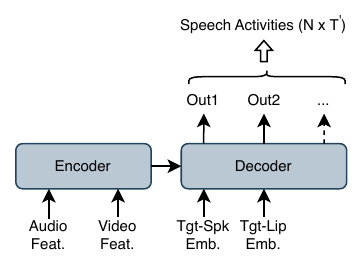}
	\label{fig_intro3}}
\caption{Overview of different TS-VAD frameworks. In (a) and (b), $T$ and $F$ denote the length and dimension of extracted audio features. $N$ and $S$ denote the number and dimension of speaker embeddings. $D$ indicates the output dimension of the decoder, which can be converted into the length of detected speech activities by a linear layer. In (c), The multi-modal encoder and multi-task decoder take multiple kinds of input (e.g., audio/video features, target-speaker/lip embeddings) and support various output methods. For clarity, front-end extractors to obtain audio-visual features are omitted in the plot.}
\label{fig_intro}
\end{figure*}

In this article, we propose a novel MIMO-TSVAD framework for audio-visual speaker diarization in modality-missing scenarios. Let $\mathbf{X}_{a}$ and $\mathbf{X}_{v}$ denote the audio and video features, respectively. $\mathbf{E}_{spk}$ represents the target-speaker embeddings in classical TS-VAD systems. We additionally define target-lip embeddings to indicate speaker identities of corresponding lip tracks, denoted as $\mathbf{E}_{lip}$. Depending on the accessibility of different data during inference, there are four cases , namely $\mathbf{X}_{a}$ vs. $\mathbf{E}_{spk}$, $\mathbf{X}_{v}$ vs. $\mathbf{E}_{lip}$, $\mathbf{X}_{a}+\mathbf{X}_{v}$ vs. $\mathbf{E}_{lip}$, $\mathbf{X}_{a}+\mathbf{X}_{v}$ vs. $\mathbf{E}_{spk}+\mathbf{E}_{lip}$. Our proposed framework can flexibly process arbitrary cases with state-of-the-art performance and strong robustness.

Fig.~\ref{fig_intro} illustrates the progress of our TS-VAD research. The classical TS-VAD systems~\cite{medennikov2020target,wang2022similarity} can be abstracted into the encoder-only methods, shown in Fig.~\ref{fig_intro1}. $\mathbf{X}_{a} \in \mathbb{R}^\mathrm{T \times F}$ is extracted from input audio by traditional methods (e.g., MFCC, Fbank) or deep neural networks. $\mathbf{E}_{spk} \in \mathbb{R}^\mathrm{N \times S}$ represents the given speaker embeddings. Each speaker embedding $\in \mathbb{R}^\mathrm{S}$ has to be repeated $T$ times and concatenated with $\mathbf{X}_{a}$, producing a 3-dimensional tensor with the shape of $T \times N \times (F+S) $. Then, encoders (e.g., Bi-LSTM, Transformers) process the input tensor along the time axis ($T$) and speaker axis ($N$). As shown in Fig.~\ref{fig_intro2}, we introduce the sequence-to-sequence architecture~\cite {cheng2023target} to factorize the input onto the encoder and decoder separately. This way, the consumption of input memory becomes proportional to $T \times F + N \times S$, enabling the model to process longer audio and more speakers. Moreover, in the previous encoder-only architecture, the output feature length ($T^{\prime}$) must be equal to the input feature length ($T$). In the sequence-to-sequence architecture, adjusting the output dimension ($T^{\prime}$) of the last linear layer can be easily implemented. As the model accepts a fixed-length speech chunk as input when $T$ is determined, using a larger output length ($T^{\prime}$) to predict the fixed-length voice activities can achieve a higher temporal resolution. In this work, the MIMO-TSVAD framework is designed based on the sequence-to-sequence architecture, demonstrated in Fig.~\ref{fig_intro3}. The multi-modal encoder and multi-task decoder leverage cross-modal and inter-speaker relationships, which can utilize various types of input data to obtain different predictions in a Multi-Input and Multi-Output (MIMO) manner.

This paper extends our previous work that presents the fundamental algorithm of Seq2Seq-TSVAD~\cite{cheng2023target}. Also, a simple version based on lip information has been initially described in our technical report~\cite{cheng2023whu} for the Multi-Modal Information based Speech Processing (MISP) 2022 Challenge~\cite{wang2023multimodal}. The new contributions from this paper are summarized as follows.

\begin{itemize}
  \item \textit{Unified}: The MIMO-TSVAD framework achieves state-of-the-art performance in the audio-based, video-based, and audio-visual speaker diarization tasks. 
  \item \textit{Flexible}: The proposed framework is jointly designed with an effective multi-stage training strategy. The integrated model can handle various kinds of input features and speaker profiles.
  \item \textit{Robust}: The proposed framework is jointly designed with an effective multi-stage inference strategy to robustly utilize audio-visual data in modality-missing scenarios.
\end{itemize}

\section{Related Works}
\label{sec:realted}

\subsection{Modularized Speaker Diarization}

The modularized speaker diarization works in a cascaded pipeline. First, a Voice Activity Detection (VAD)~\cite{chang2018temporal} module detects active speech in the audio. Next, speech regions are divided into multiple short segments through speech segmentation, such as Speaker Change Detection (SCD)~\cite{hruz2017convolutional} or uniform segmentation~\cite{sell2018diarization}. After extracting speaker representations (e.g., i-vectors~\cite{dehak2010front}, x-vectors~\cite{snyder2018x}) from those segments, a scoring metric (e.g., cosine distance, probabilistic linear discriminate analysis~\cite{sell2014speaker}) measures the pairwise embedding similarities. These segments are finally grouped into different identities by clustering algorithms such as K-Means~\cite{wang2018speaker}, Agglomerative Hierarchical Clustering (AHC)~\cite{sell2018diarization}, Spectral Clustering~\cite{lin2019lstm,lin2020self}, and so on.

The clustering module of the modularized method can flexibly estimate the number of speakers in long audio. However, it typically assumes that each audio segment contains only one speaker. To address this problem, overlapped speech detection~\cite{landini2020but} and overlap-aware resegmentation~\cite{bredin2021end,landini2022bayesian} can improve the performance by assigning multiple speaker labels to overlapped regions. Speech separation also has been adopted in offline~\cite{barker2018fifth,watanabe2020chime,xiao2021microsoft} and online~\cite{morrone2023low} diarization systems.

\subsection{End-to-End Neural Diarization}
The EEND framework~\cite{fujita2019end_1,fujita2019end_2} formulates the diarization problem as a multi-label classification task, relying on the permutation-invariant training to predict all speakers' voice activities simultaneously. The initial EEND models have a fixed number of output speakers limited by the network architecture. Although the Encoder-Decoder based Attractor (EDA)~\cite{horiguchi2020end,horiguchi2022encoder} enables EEND-based methods to process audio with a variable number of speakers, the maximum number of speakers is empirically bounded by the training data. To make the number of output speakers flexible and unlimited, EEND-vector clustering (EEND-VC)~\cite{kinoshita2021integrating,kinoshita2021advances,kinoshita2022tight} integrates end-to-end and clustering approaches, which deploys an EEND model for shortly divided audio blocks and then matches the inter-block speaker labels by clustering on speaker embeddings. In addition, EEND-GLA~\cite{horiguchi2021towards,horiguchi2022online} calculates local attractors from each short block and finds the speaker correspondence based on similarities between inter-block attractors. As its training only requires relative speaker labels within the recording, EEND-GLA is practical for adapting models on in-the-wild datasets without globally unique speaker labels.

Also, several studies promote the EEND-based systems to online inference~\cite{han2021bw,xue2021online,horiguchi2022online} or improve them in terms of advanced neural network architecture~\cite{liu2021end,rybicka2022end,fujita2023neural,chen2023attention}, objective function design~\cite{leung2021robust,jeoung2023improving}, unsupervised/semi-supervised learning~\cite{ding2020self,dissen2022self,takashima2021semi}, and so on.

\subsection{Target-Speaker Voice Activity Detection}

The background of TS-VAD can be traced back to the personal VAD~\cite{ding2019personal}, which utilizes a given acoustic footprint as the speaker profile to retrieve his or her personalized voice activities. However, the personal VAD only accepts a single speaker at once and ignores the inter-speaker modeling in a conversation. Thus, TS-VAD is designed to process multiple speaker profiles and predict their voice activities simultaneously.

The initial TS-VAD~\cite{medennikov2020target} takes speech features (e.g., MFCC) and i-vectors of each speaker as the input, where the output number of speakers is fixed. He et al.~\cite{he2021target} modify the model to cope with a flexible number of speakers by determining the maximum number of speakers and outputting null speech activities for zero-padded speaker profiles. Later, LSTM~\cite{Cheng_2023} and Transformer~\cite{wang2023target} modules are implemented along the speaker dimension of models to handle a variable number of speakers. On the other hand, the i-vectors used in TS-VAD are relatively domain-dependent, restricting the system performance on multi-scenario datasets~\cite{wang2021scenario}. This finding paves the way for exploring more discriminative speaker embeddings like x-vectors as an alternative. Wang et al.~\cite{wang2022similarity} first replace the front-end of TS-VAD with a pre-trained module tailored for extracting frame-level x-vectors. This modification shows superior robustness and generalization than a simple swap of i-vectors for x-vectors in early attempt~\cite{medennikov2020target}.

Furthermore, the TS-VAD framework has been investigated for multi-channel signal~\cite{wang2022cross}, vision-guided system~\cite{cheng2023whu}, and online inference~\cite{wang2022online,wang2023end}. Integrating features of both TS-VAD and EEND methods into an entire system has also become a popular trend~\cite{wang2022incorporating,wang2023target,chen2023attention,chen2023attention2}.

\begin{figure*}[t]
\centering
  \includegraphics[width=1.0\linewidth]{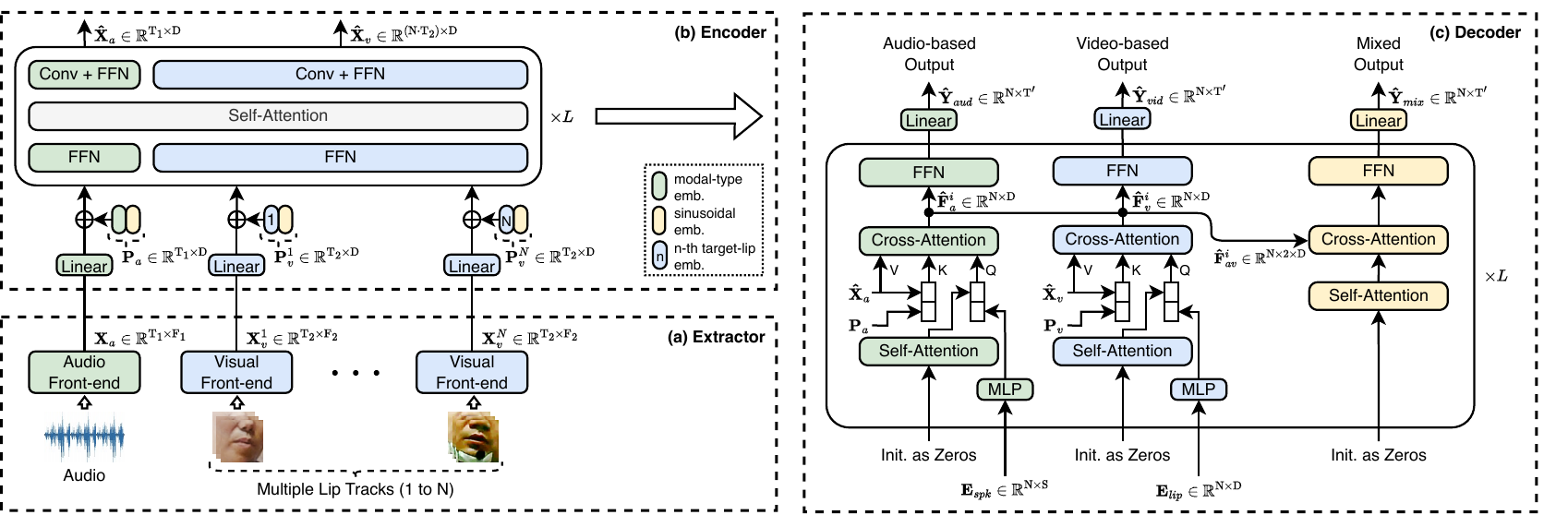}
  \caption{The MIMO-TSVAD framework. (a) Extractor: Front-end modules extract audio features $\mathbf{X}_{a}$ and video features $\mathbf{X}_{v}$ from the input data. (b): Encoder: Conformer-based modules leverage multi-modal information to generate the updated audio-visual features $\mathbf{\hat{X}}_{a}$ and $\mathbf{\hat{X}}_{v}$. (c) Decoder: Multi-task decoding modules predict voice activities based on different speaker profiles, including target-speaker embeddings $\mathbf{E}_{spk}$, target-lip embeddings $\mathbf{E}_{lip}$, or both. Green, blue, and yellow depict sub-components for processing audio, visual, and audio-visual modalities. For clarity, layer normalization and residual connection of each self-attention, cross-attention, and feedforward layer are omitted in the plot.}
  \label{fig_mimo}
\end{figure*}

\subsection{Audio-Visual Speaker Diarization}

As facial attributes and lip movements have been proven to be highly related to speech~\cite{yehia1998quantitative}, most early audio-visual speaker diarization methods leverage multi-modal cues by modeling the correspondence between speech signals and talking faces~\cite{el2014audiovisual,kapsouras2017multimodal}. Also, sound source localization using microphone arrays can establish another cross-modal relationship by mapping the speech direction onto the captured image plane~\cite{gebru2017audio,chung2019said}. However, it is sometimes hard to perfectly enroll talking faces or lip movements with related speech segments. Off-screen speakers whose faces are not captured can lead to face detection failure. This ``enroll first, diarize later'' paradigm may fail in modality-missing scenarios.

Recently, Xu et al.~\cite{xu2022ava} create the AVA Audio-Visual Diarization (AVA-AVD) database containing diverse movie clips and a multi-stage audio-visual speaker diarization system. The proposed method first utilizes VAD module to find speaking segments in the audio, then applies the ASD~\cite{tao2021someone} module to locate a face for each speaking segment. The paired audio-visual inputs are jointly scored to cluster the speaking segments into different identities. Subsequently, Wuerkaixi et al.~\cite{wuerkaixi2022dyvise} introduce the lip movement to learn a Dynamic Vision-guided Speaker Embedding (DyViSE) instead of the still facial feature, achieving new state-of-the-art performance on these real-world videos.

In addition, the MISP 2022~\cite{wang2023multimodal} database is presented for Chinese Home-TV scenarios. It has over 100 hours of audio-visual signals synchronously captured by different devices, nearly 3.5 times larger than the AVA-AVD~\cite{xu2022ava}. The MISP 2022 Challenge has been successfully held based on this database. Meanwhile, an End-to-End Audio-Visual Speaker Diarization (AVSD)~\cite{he2022end} is presented as a competitive baseline method that predicts speech probabilities using audio features and each speaker's lip video. In the AVSD system, available lip videos decide the number and order of output speakers. This way is highly effective when the camera captures all speakers ideally. In contrast, it cannot deal with the out-of-screen speakers.

Apparently, the primary ideas of most existing audio-visual speaker diarization methods originate from the previous audio-based methods. Observing the success of TS-VAD in audio-only speaker diarization and its lack of audio-visual research, this paper further extends the concept of our previous Seq2Seq-TSVAD~\cite{cheng2023target} to build a unified, flexible, and robust framework for audio-visual speaker diarization.

\section{Multi-Input Multi-Output Target-Speaker Voice Activity Detection}
\label{sec:methods}

\subsection{Architecture}

Fig.~\ref{fig_mimo} demonstrates our proposed MIMO-TSVAD framework, consisting of three parts: the extractor, the encoder, and the decoder. 

\subsubsection{Audio Extractor} 

The ResNet-34~\cite{he2016deep} is adopted as the audio front-end extractor. The audio signal is first transformed into log Mel-filterbank energies for the model to output a feature map $\in \mathbb{R}^\mathrm{C \times T_{1} \times H_{1} }$, where $C$, $T_{1}$, and $H_{1}$ denote the number of channels, temporal length, and width. Then, we implement the segmental statistical pooling (SSP)~\cite{wang2022similarity} method to aggregate channel-wise features and obtain the audio features $\mathbf{X}_{a} \in \mathbb{R}^\mathrm{T_{1} \times F_{1}}$, where $F_{1}$ is the output dimension of the SSP layer. This process can be viewed as a neural network-based feature extraction that transforms raw audio signals into frame-level representations.

\subsubsection{Video Extractor}
 
The ResNet18-3D~\cite{tran2018closer} is adopted as the video front-end extractor. A few modifications are employed based on its standard implementation in PytorchVideo\footnote{\url{https://github.com/facebookresearch/pytorchvideo}}. First, the stem layer is set to the convolutional kernel size of 7, stride of 2, and output channels of 32 without the pooling layer. The stride of pooling and convolutional layers in the residual blocks is set to $\left (1,2,2\right ) $ without the temporal downsampling. Finally, a spatial global average pooling layer is placed at the tail. The model transforms a lip video with the length of $T_{2}$ and resolution of $H \times W$ into frame-level representations $\mathbf{X}_{v} \in \mathbb{R}^\mathrm{T_{2} \times F_{2}}$, where $F_{2}$ represents the feature dimension.

\subsubsection{Encoder}

We utilize Conformer-based~\cite{gulati2020conformer} encoder to model long-term and cross-modal relationships between the extracted audio-visual representations. The layout of the designed multi-modal encoder is shown in Fig.~\ref{fig_mimo}b, which is mainly stacked by three types of basic modules: the feedforward neural networks, self-attention layers, and convolutional blocks. Inspired by the Mixture of Modality Experts (MoME) in vision-language models~\cite{bao2022vlmo}, the weights of feedforward and convolutional modules are modality-dependent, and shared self-attention layers exchange cross-modal information.

In a TSVAD-like system, the input order of speaker profiles determines the output order of voice activities. Unlike the mixed audio signals, different lip videos are naturally separate tracks to provide features as well as the role of speaker profiles. Hence, we utilize a set of learnable target-lip embeddings to indicate the relative identities of input $N$ lip tracks, denoted as $\mathbf{E}_{lip} \in \mathbb{R}^\mathrm{N \times D}$. Each element $\in \mathbb{R}^\mathrm{D}$ in $\mathbf{E}_{lip}$ is repeated to the length of $T_{2}$ and added with sinusoidal encodings~\cite{vaswani2017attention}, resulting in the positional embeddings $\mathbf{P}^{n}_{v} \in \mathbb{R}^\mathrm{T_{2}\times D}$ for the $n$-th video. The final positional embeddings for all video features can be concatenated as $\mathbf{P}_{v} \in \mathbb{R}^\mathrm{(N\cdot T_{2})\times D}$. Meanwhile, learnable modality-type embeddings $\mathbf{E}_{aud} \in \mathbb{R}^\mathrm{D}$ are initialized to differentiate the audio features from video features. Similarly, $\mathbf{E}_{aud}$ is repeated to the length of $T_{1}$ and added with sinusoidal encodings, resulting in the final positional embeddings $\mathbf{P}_{a} \in \mathbb{R}^\mathrm{T_{1} \times D}$ for audio features. Two modality-dependent linear layers map the extracted audio-visual features to the encoder dimension $D$. Then, the encoder takes the sum of audio-visual features and corresponding positional embeddings as the input. The encoded audio-visual features are denoted as $\mathbf{\hat{X}}_{a}$ and $\mathbf{\hat{X}}_{v}$, implying cross-modal information.

\subsubsection{Decoder}
In the Seq2Seq-TSVAD~\cite{cheng2023target}, the presented Speaker-Wise Decoder (SW-D) estimates target-speaker voice activities by processing encoded audio features and cross-speaker relationships simultaneously. We further extend its multi-task abilities for inference with different kinds of speaker profiles. As shown in Fig.~\ref{fig_mimo}c, the multi-task speaker-wise decoder involves three output branches. The audio-based output (green-colored) performs as same as the previous audio-only TS-VAD, utilizing target-speaker embeddings $\mathbf{E}_{spk}$ as speaker profiles to estimate multiple voice activities from the encoded audio features $\mathbf{\hat{X}}_{a}$. The video-based output (blue-colored) utilizes the newly introduced target-lip embeddings $\mathbf{E}_{lip}$ to serve as speaker profiles, detecting voice activities from the encoded video features $\mathbf{\hat{X}}_{v}$. The mixed output (yellow-colored) directly fuses intermediate features of the two modalities by cross-attention mechanism, achieving better usage of their complementary information.

The input of each decoder branch is initialized by zero embeddings $\in \mathbb{R}^\mathrm{N \times D}$ and updated step by step in the subsequent blocks, where $N$ denotes the number of candidate speakers and $D$ represents the decoder dimension. The self-attention layer is adopted to exchange inter-speaker relationships. For the audio-based and video-based branches, $\mathbf{E}_{spk}$ and $\mathbf{E}_{lip}$ go through multi-layer perception (MLP) modules to align the feature dimension with the decoder dimension $D$ and concatenate with the original queries in respective branch. The adopted MLP module consists of two linear layers with in-between layer normalization and ReLU activation. Meanwhile, we concatenate keys from $\mathbf{\hat{X}}_{a}$ and $\mathbf{\hat{X}}_{v}$ with corresponding positional embeddings $\mathbf{P_{a}}$ and $\mathbf{P_{v}}$. This way, all key-query calculations in the cross-attention layer can incorporate positional and speaker-related information as the auxiliary feature. Additionally, let $\mathbf{\hat{F}}^{i}_{a} \in \mathbb{R}^\mathrm{N \times D}$ and $\mathbf{\hat{F}}^{i}_{v} \in \mathbb{R}^\mathrm{N \times D}$ denote the intermediate features of audio-based and video-based branches in the $i$-th decoder block, respectively. Then, they are concatenated to produce the audio-visual features $\mathbf{\hat{X}}_{av} \in \mathbb{R}^\mathrm{N \times 2 \times D}$. For the mixed branch, it takes the second (modality) dimension of $\mathbf{\hat{X}}_{av}$ as the sequence axis to perform the cross-attention operation. This way, two modalities are dynamically fused into the mixed branch. Finally, a linear projection layer with sigmoid activation transforms the decoder embeddings into posterior probabilities of the estimated voice activities. The output dimension $T^{\prime}$ of the linear projection controls the temporal resolution of system prediction. For example, if the chunk length of input audio-visual data is fixed at $L_{chunk}$ seconds. The prediction for this duration will be evenly divided into $T^{\prime}$ frames. Each frame-level output indicates whether the target speaker is speaking during the corresponding time interval. In this case, the temporal resolution can be calculated as $L_{chunk}/T^{\prime}$, representing the unit duration of each frame-level prediction. By adjusting the linear projection layer, $T^{\prime}$ can be easily set as several multiples of $T_{1}$ or $T_{2}$. Given $N$ speakers, the predictions from the audio-based, video-based, and mixed output are represented as $\mathbf{\hat{Y}}_{aud}$, $\mathbf{\hat{Y}}_{vid} $, and $\mathbf{\hat{Y}}_{mix} \in \mathbb{R}^\mathrm{N \times T^{\prime}}$, respectively.

It is worth noting that the original version of SW-D in Seq2Seq-TSVAD~\cite{cheng2023target} also introduces target-speaker and positional embeddings into the self-attention layers. This operation is found to be unnecessary in this work as the decoder still works well after removing it.

\begin{figure*}[t]
\centering
  \includegraphics[width=1.0\linewidth]{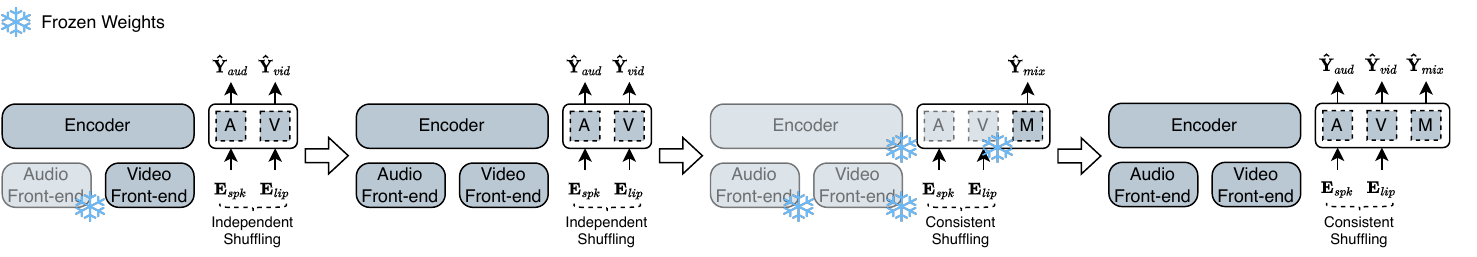}
  \caption{Multi-stage training process. $A$, $V$, and $M$ represent the decoder's audio-based, video-based, and mixed output branches. The predictions from corresponding branches are denoted as $\mathbf{\hat{Y}}_{aud}$, $\mathbf{\hat{Y}}_{vid}$, and $\mathbf{\hat{Y}}_{mix}$, respectively.}
  \label{fig_training}
\end{figure*}

\subsection{Multi-Stage Training}
\label{sec:training}

To process multiple input and output types in a unified framework, we design a multi-stage training strategy to optimize the model progressively, as shown in Fig.~\ref{fig_training}. Meanwhile, we introduce two categories of modality masking techniques during training, which enable the model to process audio-only, video-only, or audio-visual signals flexibly. The training process can be described as follows.
\begin{itemize}
  \item \textit{Stage 1}: We copy and freeze the parameters of the pre-trained speaker embedding model to initialize the audio front-end extractor. Only audio and video-based output branches are adopted to train on fully simulated data.  
\item \textit{Stage 2}: The audio front-end model is unfrozen. Training data for model adaption on the specific dataset is added at a given ratio.
  \item \textit{Stage 3}: All pre-trained parameters are frozen. The mixed output branch is initialized and trained separately. 
  \item \textit{Stage 4}: Finally, all parameters are unfrozen to be finetuned jointly.
  \end{itemize}

\begin{figure}[t]
\centering
  \includegraphics[width=1.0\linewidth]{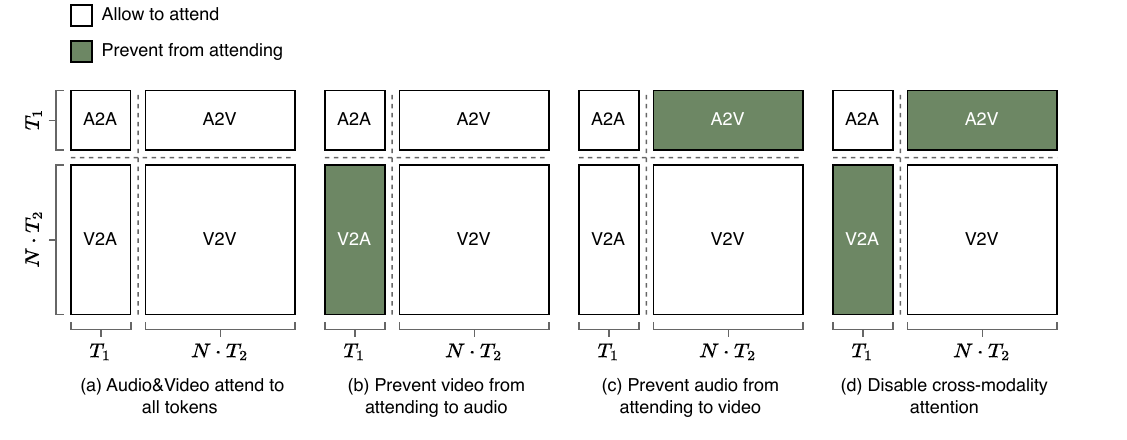}
  \caption{Different attention masks in the encoder. The first $T_{1}$ tokens of the input feature sequence are the audio features. The rest $N\cdot T_{2}$ tokens are video features extracted from $N$ lip tracks.}
  \label{fig_mask}
\end{figure}

\subsubsection{Model-Level Modality Masking}

In training \textit{stages 1-2}, the mixed output branch in the decoder is not initialized. Cross-modal information is only exchanged by self-attention layers in the encoder. Hence, we implement different attention masks to adapt the model to different input features. Fig.~\ref{fig_mask} demonstrates four possible cases, reflecting that the information can be bidirectional, one-way, or prohibited from flowing between two modalities. During training, one of the masks is randomly selected. During inference, the attention mask is generated according to the existence of each modality. In other words, tokens should only attend to the existing modalities. To decouple two output branches to avoid learning shortcuts from matched target-speaker and target-lip embeddings, the speaker order of $\mathbf{E}_{lip}$ should be shuffled independently from the $\mathbf{E}_{spk}$. As $\mathbf{E}_{lip}$ represents the relative identities, shuffling $\mathbf{E}_{lip}$ can be done equivalently by shuffling the input order of lip tracks in practice. Ground truth labels for different output branches must be re-assigned based on their shuffled results.  

Given an audio-visual recording with $N$ existing speakers, except target-lip embeddings $\mathbf{E}_{lip}$ as learnable parameters built in the model, the other inputs involve audio signal $\mathbf{A}$, target-speaker embeddings $\left \{\mathbf{e}_{n} \mid 1 \le n \le N  \right \}$, and lip tracks $\left \{\mathbf{l}_{n} \mid 1 \le n \le N  \right \}$. Each $\mathbf{e}_{n}$ and $\mathbf{l}_{n}$ represents the $n$-th speaker embedding and lip track, respectively. The ground truth labels for voice activities can be denoted as a binary matrix $\mathbf{Y} \in \left ( 0,1 \right )^\mathrm{N \times T^{\prime}}$, where $\mathbf{Y}(n,t)$ represents the speaking existence of the $n$-th speaker at time $t$. The audio-based output $\mathbf{\hat{Y}}_{aud}$ and video-based output $\mathbf{\hat{Y}}_{vid}$ are modeled with our proposed MIMO-TSVAD as follows:
\begin{align}
& \mathbf{E} = sh_{1}\left ( \left \{\mathbf{e}_{n} \mid 1 \le n \le N  \right \} \right ), \\
& \mathbf{L} = sh_{2}\left ( \left \{\mathbf{l}_{n} \mid 1 \le n \le N  \right \} \right ),  \\
& \mathbf{\hat{Y}}_{aud}, \mathbf{\hat{Y}}_{vid} = \text{MIMO\_TSVAD} \left (\mathbf{A}, \mathbf{E}, \mathbf{L}  \right ),
\end{align}	

\noindent where $sh_{1}\left (\cdot \right ) $ and $sh_{2}\left (\cdot \right ) $ represent two independent operations for shuffling speaker-orders ($n$) in $\mathbf{E}$ and $\mathbf{L}$, respectively. The audio-based output loss $\mathcal{L}_{aud}$ and video-based output loss $\mathcal{L}_{vid}$ are described as:
\begin{align}
& \mathcal{L}_{aud} = BCE\left (sh_{1}\left (\mathbf{Y} \right ), \mathbf{\hat{Y}}_{aud} \right ), \\
& \mathcal{L}_{vid} = BCE\left (sh_{2}\left (\mathbf{Y} \right ), \mathbf{\hat{Y}}_{vid} \right ),
\end{align}	

\noindent where ground truth labels $\mathbf{Y}$ are re-assigned based on the same shuffling operations. $BCE\left (y, \hat{y} \right )$  measures the binary cross-entropy between the target $y$ and predicted $\hat{y}$. This way, the trained encoder can handle different input features.

\subsubsection{Data-Level Modality Masking}

In training \textit{stages 3-4}, the mixed output branch in the decoder is additionally introduced to combine the audio-based and video-based output, which requires that speaker orders of $\mathbf{E}_{spk}$ and $\mathbf{E}_{lip}$ are consistent. However, in realistic scenarios, it is difficult for each speaker to obtain perfectly paired target-speaker embedding and lip movement all the time. Hence, we implement different data masks to adapt the model to uncertain speaker profiles. During training, each speaker has a probability of 0.5 to conduct data masking. Once the $n$-th speaker is selected, either the speaker embedding $\mathbf{e}_{n}$ or lip track $\mathbf{l}_{n}$ will be masked by zeros. Let $\mathbf{M}_{spk}^{in} = \left \{a_{n} \mid a_{n} \in \{0, 1\}, 1 \le n \le N  \right \}$ and $\mathbf{M}_{spk}^{out} \in \left ( 0,1 \right )^\mathrm{N \times T^{\prime}}$ record the masking states of target-speaker embeddings. If the $n$-th speaker embedding is masked, $a_{n} \in \mathbf{M}_{spk}^{in}$ and $\mathbf{M}_{spk}^{out}(n,:)$ will be set to zeros. Similarly, the masking states of lip tracks are denoted as $\mathbf{M}_{lip}^{in} = \left \{b_{n} \mid b_{n} \in \{0, 1\}, 1 \le n \le N  \right \}$ and $\mathbf{M}_{lip}^{out} \in \left ( 0,1 \right )^\mathrm{N \times T^{\prime}}$. If the $n$-th lip track is masked, $b_{n} \in \mathbf{M}_{lip}^{in}$ and $\mathbf{M}_{lip}^{out}(n,:)$ will be set to zeros. In this situation, the proposed model output can be obtained as follows:

\begin{align}
& \mathbf{E}^{\prime} = sh\left ( \left \{ a_{n} \times \mathbf{e}_{n} \mid 1 \le n \le N  \right \} \right ), \\
& \mathbf{L}^{\prime} = sh\left ( \left \{ b_{n} \times \mathbf{l}_{n} \mid 1 \le n \le N  \right \} \right ),  \\
& \mathbf{\hat{Y}}_{aud}^{\prime}, \mathbf{\hat{Y}}_{vid}^{\prime}, \mathbf{\hat{Y}}_{mix}^{\prime} = \text{MIMO\_TSVAD} \left (\mathbf{A}, \mathbf{E}^{\prime}, \mathbf{L}^{\prime}  \right ),
\end{align}	

\noindent where $sh\left (\cdot \right )$ represents the consistent speaker-order shuffling operation. Each input $\mathbf{e}_{n}$ and  $\mathbf{l}_{n}$ will be retained or zeroed by multiplying by the $a_{n} \in \{0, 1\}$ and $b_{n} \in \{0, 1\}$. The audio-based and video-based output losses are modified as follows: 

\begin{align}
& \mathcal{L}_{aud}^{\prime} = BCE\left (sh(\mathbf{M}_{spk}^{out}\wedge \mathbf{Y}), \mathbf{\hat{Y}}_{aud}^{\prime} \right ), \\
& \mathcal{L}_{vid}^{\prime} = BCE\left (sh(\mathbf{M}_{lip}^{out}\wedge \mathbf{Y}), \mathbf{\hat{Y}}_{vid}^{\prime} \right ), 
\end{align}

\noindent where $\wedge$ denotes the element-wise logical AND operation. Ground truth labels $\mathbf{Y}$ are not only re-assigned by speaker-order shuffling but also filtered by the masking states. Then, the mixed output loss $\mathcal{L}_{mix}^{\prime}$ is written by:

\begin{align}
\mathcal{L}_{mix}^{\prime} &= BCE\left (sh(\left (\mathbf{M}_{spk}^{out} \vee \mathbf{M}_{lip}^{out} \right )\wedge \mathbf{Y}), \mathbf{\hat{Y}}_{mix}^{\prime} \right ),
\end{align}

\noindent where $\vee$ denotes the element-wise logical OR operation. 

Since the masking probability is independent for all speakers, speaker embeddings or lip tracks of multiple speakers may be masked at a time. Because for the same speaker, we can only randomly mask either the speaker embedding or lip track simultaneously. If the $n$-th speaker is masked among all $N$ inputs, the model still generates an output of shape $N \times T^{\prime}$. As long as one of the $\mathbf{e}_{n}$ or $\mathbf{l}_{n}$ is available, the mixed output branch of the proposed model can work properly for the $n$-th speaker. This way, the trained decoder can be compatible with uncertain speaker profiles.

To summarize, the first two stages utilize the model-level modality masking to address different input features (e.g., audio-only, video-only, or audio-visual), which are optimized by the total diarization loss of $\mathcal{L}_{aud}+\mathcal{L}_{vid}$. Meanwhile, it prevents the model from receiving many zero-masked features caused by the data-level modality masking in early training. The last two stages utilize the data-level modality masking to solve uncertain inference conditions with modality-mismatched speaker profiles (e.g., target-speaker embeddings, target-lip embeddings, or mixed), which are optimized by the total diarization loss of $\mathcal{L}_{aud}^{\prime}+\mathcal{L}_{vid}^{\prime}+\mathcal{L}_{mix}^{\prime}$. Finally, the trained model can flexibly deal with varying accessibilities of audio-visual data.

\begin{figure*}[t]
\centering
  \includegraphics[width=1.0\linewidth]{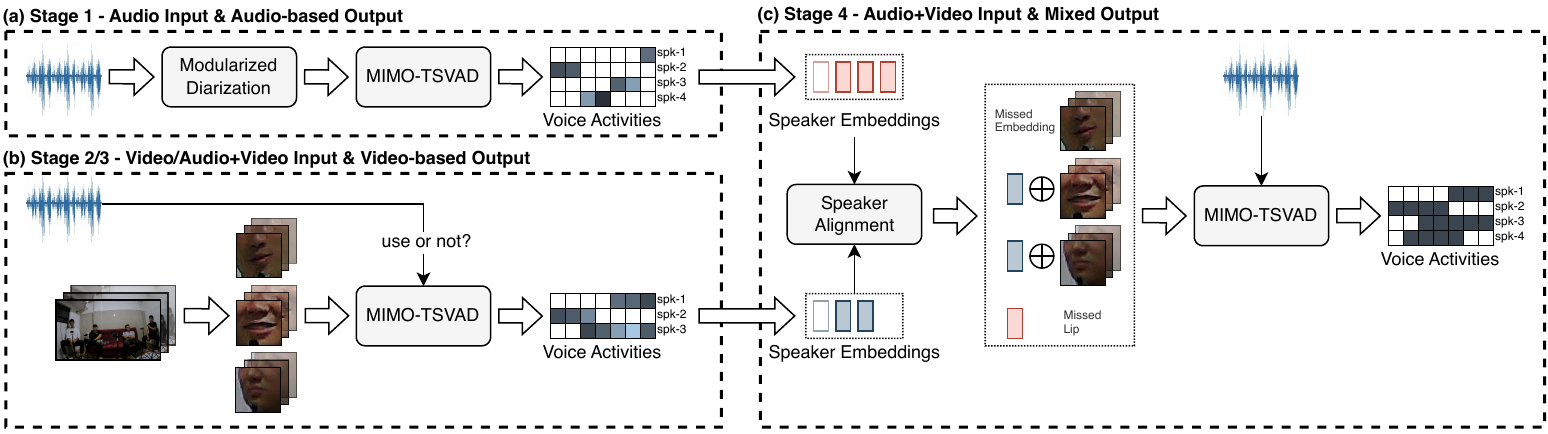}
  \caption{Multi-stage inference process. The MIMO-TSVAD model is weight-shared in all stages. (a) The MIMO-TSVAD model is a typical audio-based system in inference \textit{stage 1}. (b) The MIMO-TSVAD decoder is based on the video-based output. When the encoder takes video-only input, it operates in inference \textit{stage 2}. Otherwise, it operates in inference \textit{stage 3} if the audio input is also available. (c) The speaker alignment module finds off-screen speaker embedding to build the new set of mixed speaker profiles. Then, the MIMO-TSVAD model operates in inference \textit{stage 4} to take audio-visual input and mixed output. In this example, the first speaker embedding is not successfully extracted because his non-overlapped speech duration is too short. Also, the last speaker's lip track is undetected, but the speaker alignment module replenishes his speaker embedding (pink-colored).}
  \label{fig_inference}
\end{figure*}

\subsection{Multi-Stage Inference}
\label{sec:inference}

Given the synchronized audio-visual data, the algorithm aims to find all speaker identities and localize their speaking regions automatically. To adequately utilize audio-visual data in modality-missing scenarios, we design a multi-stage inference strategy to predict voice activities iteratively.

\subsubsection{Prior Steps}

Following the commonly used paradigm in previous TS-VAD methods~\cite{medennikov2020target,wang2022similarity,cheng2023target,wang2023target}, a modularized diarization system is necessary to obtain an initial result. Each detected speaker's non-overlapped speech segments are aggregated to extract the target embedding, initializing the MIMO-TSVAD model to conduct audio-based speaker diarization. 

For the video signals, we extract each speaker's lip track as the same as our previous works~\cite{cheng2023whu,cheng2022dku}. The RetinaFace~\cite{deng2019retinaface} detector is deployed to localize face images with five-point facial landmarks in each video frame. As a talking face is assumed not to move dramatically in a short time window, the K-Means algorithm in Scikit-Learn toolkit~\cite{pedregosa2011scikit} utilizes coordinates of detected faces to cluster the same person in adjacent frames. After obtaining each speaker's face track, we crop the lip region of interest (Lip-RoI) based on an empirical setting reported in the CAS-VSR-W1k~\cite{yang2019lrw} database for large-scale lip reading tasks. Let $\mathbf{p_{1}}$, $\mathbf{p_{2}}$, and $\mathbf{p_{3}}$ represent coordinates of the nose tip, left mouth corner, and right mouth corner. The Lip-RoI bounding box is defined as:
\begin{align}
x_{center},y_{center} = \frac{\mathbf{p_{2}} + \mathbf{p_{3}}}{2},
\end{align}
\begin{align}
width = min\left \{3.2\times d_{MN} , 2\times max \left \{d_{MN}, d_{p2p3} \right \}  \right \}, 
\end{align}

\noindent where $d_{MN}$ denotes the Euclidean distance between the nose tip and the center of the mouth, $d_{p2p3}$ denotes the Euclidean distance between $\mathbf{p_{2}}$ and $\mathbf{p_{3}}$. Furthermore, undetected frames of each lip track will be padded by zeros, guaranteeing audio-visual synchronization. As this work mainly focuses on the TS-VAD research, we adopt this simple detection and tracking method to extract multiple speakers' lip tracks.

\subsubsection{Diarization}

The MIMO-TSVAD framework requires at least two input parts: features and speaker profiles. The features (e.g., audio, video, or audio-visual data) imply rich voice activity information in conversations. The speaker profiles (e.g., target-speaker embeddings, target-lip embeddings) are the reference to separate individual speakers' corresponding voice activities from the provided features. As mentioned in Section~\ref{sec:intro}, the MIMO-TSVAD framework supports four cases according to the accessibility of different input data, which can be described as a multi-stage inference process.

\begin{itemize}
  \item \textit{Stage 1}: The model takes audio features $\mathbf{X}_{a}$ and target-speaker embeddings $\mathbf{E}_{spk}$, which is an audio-only system.
  \item \textit{Stage 2}: The model takes video features $\mathbf{X}_{v}$ and target-lip embeddings $\mathbf{E}_{lip}$, which is a video-only system.  
  \item \textit{Stage 3}: The model takes audio-visual features $\mathbf{X}_{a}+\mathbf{X}_{v}$ and target-lip embeddings $\mathbf{E}_{lip}$. The only difference with the \textit{stage 2} is using auxiliary audio features.
  \item \textit{Stage 4}: The model takes audio-visual features $\mathbf{X}_{a}+\mathbf{X}_{v}$ and the mix of target-speaker and target-lip embeddings $\mathbf{E}_{spk}+\mathbf{E}_{lip}$. Each speaker's target speech activities can be extracted if at least one kind of speaker profile exists.
\end{itemize}

Fig.~\ref{fig_inference} illustrates the overall inference process for an audio-visual recording. First, the audio signal undergoes the modularized diarization system to obtain initial speaker profiles. Then, the inference \textit{stage 1} of MIMO-TSVAD predicts voice activities based on the audio-based output. This method ensures an audio-based solution is still available even if all visual information is lost. For lip tracks extracted from the video signal, the inference \textit{stage 2} of MIMO-TSVAD predicts voice activities based on the video-based output. If the audio signal is fed as an auxiliary feature, this stage becomes the inference \textit{stage 3}. The inference \textit{stages 2-3} of MIMO-TSVAD directly utilizes the input lip videos to serve as speaker profiles, which does not require an additional module like the modularized method in inference \textit{stage 1}. However, a critical problem is that they can not detect the existence of completely off-screen speakers. To solve the problem, a Speaker Alignment (SA) module is deployed to find the undetected off-screen speaker embeddings. We first extract speaker embeddings based on the predicted voice activities from the inference \textit{stage 1} and \textit{stage 3}, respectively. By pairwise similarity measurement between two embedding sets, the Hungarian algorithm~\cite{kuhn1955hungarian} is employed to obtain matching relationships with the highest cosine scores. The off-screen speaker embeddings from the audio-based output can be discovered if they are not successfully matched with a pairwise similarity score higher than the pre-set speaker verification threshold. Then, the matched speaker embeddings are averaged, and the unmatched off-screen speaker embeddings are replenished. Finally, each available speaker profile can be unimodal or bimodal. The inference \textit{stage 4} of MIMO-TSVAD predicts voice activities based on the mixed output, tackling complex modality-missing situations.

During the whole inference, the pre-trained MIMO-TSVAD model is able to adopt shared weights for conducting different inference stages flexibly. All available data can be fully exploited to boost the final diarization performance, whether viewed from the aspects of audio-visual features or complementary speaker profiles.

\section{Experimental Settings}
\label{sec:exp1}

\subsection{Datasets}

The proposed MIMO-TSVAD framework is a multi-modal extension of our previous work~\cite{cheng2023target}, which supports audio-based, video-based, and audio-visual speaker diarization. We first verify its advantages on the VoxConverse~\cite{chung2020spot} and DIHARD-III~\cite{ryant2020third} datasets for audio-based speaker diarization. The VoxConverse dataset is an in-the-wild dataset with 20.29 hours of development set and 43.53 hours of test set collected from YouTube. We select the first 172 recordings (80\%) of the original development set for training MIMO-TSVAD models. The last 44 recordings (20\%) remain for validation. The DIHARD-III dataset is a multi-domain dataset with 34.15 hours of development set and 33.01 hours of evaluation set, including 11 complex scenarios (e.g., interview, clinical, restaurant). We select the first 203 recordings (80\%) of the original development set for training MIMO-TSVAD models. The last 51 recordings (20\%) remain for validation. 

In addition, the MISP 2022~\cite{wang2023multimodal} dataset is adopted for audio-visual speaker diarization. The MISP 2022 dataset targets the Chinese home-TV scenario with 106.09 hours of training set, 2.6 hours of development set, and 3.12 hours of evaluation set. Audio-visual signals are synchronously captured by different devices. The single-channel microphone records the near-field ($<0.5m$) audio. The 2-channel microphone array and high-definition camera capture the middle-field ($1-1.5m$) data. The 6-channel microphone array and wide-angle camera capture the far-field ($3-5m$) data. Also, we utilize the NARA-WPE~\footnote{\url{https://github.com/fgnt/nara_wpe}} and SETK~\footnote{\url{https://github.com/funcwj/setk}} toolkits to do dereverberation and MVDR-based beamforming for the multi-channel audio signals, augmenting the diversity of training data. Under the MISP challenge rules, all fields can be used in model training and validation. Only far-field data is allowed for evaluation.

We also create the simulated audio-visual dataset to benefit neural networks from large-scale training data. The VoxCeleb2~\cite{Chung18b} dataset is adopted as the source audio corpus, which covers thousands of unique speaker identities. As the video contents of the VoxCeleb2 dataset are not available for downloading nowadays, we utilize the MISP 2022~\cite{wang2023multimodal} dataset as a supplementary video corpus. Although the audio and video signals are not semantically consistent regarding the dialogue content, the combination is still feasible because speaker diarization primarily focuses on recognizing speaker identities and voice activities rather than the speech content. During training, each simulated data is generated in an on-the-fly manner as follows.
\begin{itemize}
  \item \textit{Step 1}: A speaker label is randomly selected from the audio corpus. The single-speaker utterance is created by alternately concatenating his or her source speech and silent (zero-padded) segments, where each segment length is sampled from a uniform distribution of 0-4 seconds.
  \item \textit{Step 2}: We cut the video corpus into active (speech) and inactive (silent) lip tracks. An active lip track is randomly selected and cropped into the same duration for each speech segment created in the first step. An inactive lip track is randomly selected and cropped into the same duration for each silent segment. Then, the extracted lip tracks are concatenated in the same order as the audio, creating the corresponding single-speaker lip track.
  \item \textit{Step 3}: For each simulation, audio-visual data of 1-4 speakers can be obtained by repeating the above steps. All single-speaker utterances are averaged as the mixed audio. No additional operations are required for the lip tracks, as they are already separate. The average overlap ratio of simulated data is estimated to be around 30\%.
\end{itemize}

\subsection{Network Configuration}

\subsubsection{Speaker Embedding Extraction}
\label{sec:embedding}

The ResNet-34 is utilized as the pattern extractor, where residual blocks have widths (number of channels) of $\left\{64,128,256,512\right\}$ with a total downsampling factor of 8. After adding the statistical pooling~\cite{snyder2018x} layer, a linear projection layer outputs the 256-dim speaker embedding. The ArcFace ($s=32,m=0.2$)~\cite{deng2019arcface} is used as the classifier. The model trained on the VoxCeleb2~\cite{Chung18b} dataset achieves an equal error rate (EER) of 0.814\% on the Vox-O~\cite{Nagrani17} trial. Other training details are the same as~\cite{qin2022simple}. The pre-trained model is used for speaker embedding extraction in the subsequent MIMO-TSVAD system. Also, the Speaker Alignment (SA) module employs the pre-trained speaker verification threshold of 0.3479 during inference.

\begin{table*}[t]
	\centering
	\setlength{\tabcolsep}{6.5pt}
	\caption{Performance of different MIMO-TSVAD models on the VoxConverse test set (collar = 250 ms). The Diarization Error Rate (DER) is the sum of miss (MI), false alarm (FA), and speaker confusion (SC) rates.}
	\label{exp:voxconverse}
	\begin{tabular}{llllrrrrrrrrrrrr}
		\toprule
		\multirow{2}{*}{\textbf{ID}} 
		&\multirow{2}{*}{\textbf{\makecell[l]{VAD\\Resolution}}} 
		&\multirow{2}{*}{\textbf{\makecell[l]{SPK\\Capacity}}} 
		&\multirow{2}{*}{\textbf{\makecell[l]{Chunk\\Length/Shift}}} 
		&\multicolumn{4}{c}{\textbf{1-10 SPKs (\%)}} 
		&\multicolumn{4}{c}{\textbf{10+ SPKs (\%)}}
		&\multicolumn{4}{c}{\textbf{Total (\%)}}\\
		\cmidrule(lr){5-8} 
		\cmidrule(lr){9-12}
		\cmidrule(lr){13-16}
		& & & 
		&\textbf{MI} & \textbf{ FA} & \textbf{ SC} & \textbf{DER} 
		&\textbf{MI} & \textbf{ FA} & \textbf{ SC} & \textbf{DER}
		&\textbf{MI} & \textbf{ FA} & \textbf{ SC} & \textbf{DER}  \\
		\midrule
		S1 & \multirow{3}{*}{80ms} & 10 & \multirow{3}{*}{8s / 8s}
		& 2.43 & 1.48 & 1.10 & 5.01 
		& 2.80 & 10.24 & 3.44 & 16.48
		& 2.51 & 3.27 & 1.58 & 7.36
		\\ 
		S2 & & 20 & 
		& 1.99 & 1.35 & 1.05 & 4.39 
		& 2.10 & 2.52 & 3.57 & 8.19 
		& 2.01 & 1.59 & 1.56 & 5.16
		\\
		S3 & & 30 & 
		& 2.05 & 1.29 & 1.00 & 4.34 
		& 2.20 & 1.74 & 3.56 & 7.50
		& 2.08 & 1.38 & 1.52 & \textbf{4.98}
		\\
		\midrule
		S4 & \multirow{3}{*}{10ms} & 10 & \multirow{3}{*}{8s / 8s} 
		& 2.27 & 1.48 & 1.00 & 4.75
		& 2.73 & 10.08 & 3.36& 16.17
		& 2.36 & 3.24 & 1.49 & 7.09
		\\
		S5 & & 20 & 
		& 1.93 & 1.32 & 1.02 & 4.27 
		& 2.12 & 2.41 & 3.53 & 8.06
		& 1.97 & 1.54 & 1.54 & 5.05
		\\
		S6 & & 30 & 
		& 2.02 & 1.34 & 1.01 & 4.37
		& 2.17 & 1.72 & 3.60 & 7.49
		& 2.05 & 1.42 & 1.55 & \textbf{5.02}
		\\
		\midrule
		S7 & \multirow{3}{*}{10ms} & \multirow{3}{*}{30} & 16s / 16s
		& 2.05 & 1.20 & 0.96 & 4.21
		& 2.26 & 1.82 & 3.48 & 7.56
		& 2.09 & 1.33 & 1.48 & 4.90
		\\
		S8 &  &  & 32s / 32s
		& 2.04 & 1.16 & 0.91 & 4.11
		& 2.35 & 1.67 & 3.34 & 7.36
		& 2.10 & 1.26 & 1.40 & 4.76
		\\
		S9 &  &  & 64s / 64s
		& 1.87 & 1.15 & 0.86 & 3.88
		& 2.26 & 1.66 & 3.15 & 7.07
		& 1.94 & 1.26 & 1.33 & \textbf{4.53}
		\\
		\midrule
		S10 & \multirow{3}{*}{10ms} & \multirow{3}{*}{30} & 16s / 2s
		& 1.95 & 1.02 & 0.85 & 3.82
		& 2.22 & 1.39 & 3.28 & 6.89
		& 2.01 & 1.10 & 1.35 & 4.46
		\\
		S11 &  &  & 32s / 2s
		& 1.95 & 0.99 & 0.79 & 3.73
		& 2.28 & 1.26 & 3.15 & 6.69
		& 2.01 & 1.05 & 1.27 & 4.33
		\\
		S12 &  &  & 64s / 2s
		& 1.85 & 1.02 & 0.74 & 3.61
		& 2.21 & 1.30 & 2.96 & 6.47
		& 1.92 & 1.07 & 1.19 & \textbf{4.18}
		\\
		\bottomrule
	\end{tabular}
\end{table*}

\subsubsection{MIMO-TSVAD}

The audio front-end ResNet-34 is initialized by the pre-trained speaker embedding model. We replace the original statistical pooling (SP) with SSP~\cite{wang2022similarity} layer to achieve frame-level feature extraction. The video front-end ResNet18-3D has residual blocks with widths of $\left\{32,64,128,256\right\}$ to output 256-dim frame-level features, which are randomly initialized. The back-end encoder-decoder modules have 6 blocks sharing the same settings: 512-dim attentions with 8 heads and 1024-dim feed-forward layers with a dropout rate of 0.1. The kernel size of convolutions in Conformer blocks is 15, and the other implementation details are the same as~\cite{gulati2020conformer}.

\subsection{Training and Inference Details}

All training data is split into fixed-length chunks by sliding window and normalized with a mean of 0 and a standard deviation of 1. The chunk length can be set to different values (e.g., 8 seconds, 16 seconds, 32 seconds). As the model only accepts a fixed-length chunk as input, the chunk lengths used for training and inference should always be consistent, but the chunk shift can vary flexibly. The subsequent experimental results investigate the impacts of different chunk lengths and shifts. Then, the input acoustic features are 80-dim log Mel-filterbank energies with a frame length of 25 ms and a frameshift of 10 ms. The input lip videos are transformed into grayscale with a resolution of $88\times88$ and frames per second (FPS) of 25. Assume that the maximum number of speaker profiles (speaker capacity) is set to $C$. When speakers in a recording cannot reach $C$, empty speaker profiles are padded by zeros or speakers not appearing in the current chunk. This padding method aligns the dimension of batched training data and forces the model to distinguish valid and invalid speaker profiles. Meanwhile, all the input speaker profiles should be shuffled to make the model invariant to speaker order.

We implement the BCE loss and Adam optimizer to train the neural network. As described in Section~\ref{sec:training}, the first training stage starts with a linear learning rate warm-up from 0 to \textit{1e-4} in 2,000 iterations. From the second training stage, real data from the VoxConverse~\cite{chung2020spot} and DIHARD-III~\cite{ryant2020third} datasets is added to the simulated data at a ratio of 0.2. For experiments on the MISP 2022~\cite{wang2023multimodal} dataset, this ratio is adjusted to 0.5, which follows our previous challenge setting~\cite{cheng2023whu} to mitigate the large domain gap between simulated data (English) and test set (Chinese). In the last finetuning stage, the learning rate is decayed to \textit{1e-5}. Additive noise from Musan~\cite{snyder2015musan} and reverberation from RIRs~\cite{ko2017study} are applied for audio augmentation. For video augmentation, input lip videos undergo each item of the following procedures with a probability of 0.5: rotation with an angle range $[5,20]$; horizontal flipping; cropping with the scale range $[0.8,1]$; transformation of contrast, brightness, and saturation in the range $[-25,25]$. The training process takes around 200k iterations with a batch size of 32 on 8 $\times$ NVIDIA RTX-3090 GPUs.

\begin{table}[t]
	\centering
	\setlength{\tabcolsep}{18pt}
	\caption{Comparisons of MIMO-TSVAD models with others on the VoxConverse test set (Collar = 250 ms).}
	\label{voxconverse_comp}
	\begin{threeparttable}[b]
	\begin{tabular}{lr}
		\toprule
		\textbf{Method} & \textbf{DER (\%)} \\
		\midrule
		ByteDance~\cite{wang2021bytedance}~\tnote{$\dag$} & 5.17 \\
		DKU-DukeECE~\cite{wang2022dku}~\tnote{$\ddag$} & 4.94  \\
		Microsoft~\cite{wang2023target} & 4.57 \\ 
		PET-TSVAD~\cite{wang2023profile}  & 4.35 \\  
		pyannote.audio~\cite{baroudi2023pyannote}~\tnote{$\wr$} & 4.00\\
		\midrule
		LSTM-SC~\cite{wang2022similarity} & 6.33 \\
		VBx~\cite{landini2022bayesian} & 5.62 \\
		AHC~\cite{wang2022dku} & 5.35 \\
		\quad $+$ MIMO-TSVAD (S12 in Table~\ref{exp:voxconverse})~\tnote{$\ast$}& \textbf{4.18} \\
		\bottomrule
	\end{tabular}
	\begin{tablenotes}
    	\item[$\dag$] VoxSRC-21 2nd-ranked result with 3-system fusion. The 1st-ranked team dose does not report the DER for this set. 
    	\item[$\ddag$] VoxSRC-22 1st-ranked result with 4-system fusion.
    	\item[$\wr$] VoxSRC-23 2nd-ranked result with WavLM~\cite{chen2022wavlm} model.
    	\item[$\ast$] participates in the VoxSRC-23 1st-ranked winning (fusion) system. 
    \end{tablenotes}
	\end{threeparttable}
\end{table}

The inference process follows the Section~\ref{sec:inference}. The prior steps extract lip tracks and target embeddings of speakers with non-overlapped speech longer than 2 seconds. If the speakers are less than speaker capacity $C$, empty speaker profiles are padded by zeros. Otherwise, the excess speaker profiles are inferred in the next group. We split test data into chunks to feed each MIMO-TSVAD inference stage. The predictions are stitched chunk by chunk. For the far-field recordings in the MISP 2022 dataset, we directly average the predictions from all dereverberated channels. Lastly, reference VAD can revise the diarization results if the specific evaluation metric allows. The timestamps marked as active speech will assign a positive label to the speaker with the highest predicted score. The predictions at non-speech timestamps will be zeroed. All experiments are repeated three times to report the mean values.

\section{Results and Discussions}
\label{sec:exp2}

\subsection{Evaluation of Audio-based Diarization}

\begin{table*}[t]
	\centering
	\setlength{\tabcolsep}{6.5pt}
	\caption{Performance of different MIMO-TSVAD models on the DIHARD-III evaluation set (Reference VAD). The Diarization Error Rate (DER) is the sum of miss (MI), false alarm (FA), and speaker confusion (SC) rates.}
	\label{exp:dihard}
	\begin{tabular}{llllrrrrrrrrrrrr}
		\toprule
		\multirow{2}{*}{\textbf{ID}} 
		&\multirow{2}{*}{\textbf{\makecell[l]{VAD\\Resolution}}} 
		&\multirow{2}{*}{\textbf{\makecell[l]{SPK\\Capacity}}} 
		&\multirow{2}{*}{\textbf{\makecell[l]{Chunk\\Length/Shift}}} 
		&\multicolumn{4}{c}{\textbf{1-5 SPKs (\%)}} 
		&\multicolumn{4}{c}{\textbf{5+ SPKs (\%)}}
		&\multicolumn{4}{c}{\textbf{Total (\%)}}\\
		\cmidrule(lr){5-8} 
		\cmidrule(lr){9-12}
		\cmidrule(lr){13-16}
		& & & 
		&\textbf{MI} & \textbf{FA} & \textbf{SC} & \textbf{DER}  
		&\textbf{MI} & \textbf{FA} & \textbf{SC} & \textbf{DER}
		&\textbf{MI} & \textbf{FA} & \textbf{SC} & \textbf{DER} \\
		\midrule
		S1 & \multirow{3}{*}{80ms} & 5 & \multirow{3}{*}{8s / 8s}
		& 4.58 & 3.17 & 3.21 & 10.96 
		& 10.92 & 2.80 & 9.56 & 23.28 
		& 5.47 & 3.11 & 4.10 & 12.68
		\\ 
		S2 & & 10 & 
		& 4.61 & 3.08 & 3.17 & 10.86
		& 10.58 & 2.58 & 9.23 & 22.39 
		& 5.45 & 3.01 & 4.03 & \textbf{12.49}
		\\
		S3 & & 20 & 
		& 4.60 & 3.19 & 3.14 & 10.93
		& 10.70 & 2.47 & 9.04 & 22.21  
		& 5.45 & 3.09 & 3.97 & 12.51
		\\
		\midrule
		S4 & \multirow{3}{*}{10ms} & 5 & \multirow{3}{*}{8s / 8s} 
		& 3.80 & 2.39 & 3.30 & 9.49
		& 10.17 & 2.74 & 9.87 & 22.78 
		& 4.69 & 2.44 & 4.22 & 11.35
		\\
		S5 & & 10 & 
		& 3.76 & 2.39 & 3.22 & 9.37
		& 10.29 & 2.02 & 9.40 & 21.71
		& 4.68 & 2.34 & 4.09 & 11.11
		\\
		S6 & & 20 & 
		& 3.94 & 2.23 & 3.19 & 9.36
		& 10.49 & 1.83 & 9.19 & 21.51
		& 4.86 & 2.18 & 4.04 & \textbf{11.08}
		\\
		\midrule
		S7 & \multirow{3}{*}{10ms} & \multirow{3}{*}{20} & 16s / 16s
		& 3.64 & 2.42 & 3.07 & 9.13
		& 10.21 & 1.88 & 8.83 & 20.92
		& 4.56 & 2.35 & 3.88 & 10.79
		\\
		S8 &  &  & 32s / 32s
		& 3.55 & 2.39 & 2.87 & 8.81
		& 9.99 & 1.88 & 8.86 & 20.73
		& 4.45 & 2.31 & 3.71 & \textbf{10.47}
		\\
		S9 &  &  & 64s / 64s
		& 3.49 & 2.63 & 2.89 & 9.01
		& 9.57 & 2.25 & 8.91 & 20.73
		& 4.34 & 2.58 & 3.73 & 10.65
		\\
		\midrule
		S10 & \multirow{3}{*}{10ms} & \multirow{3}{*}{20} & 16s / 2s
		& 3.60 & 2.23 & 2.88 & 8.71
		& 10.32 & 1.59 & 8.35 & 20.26
		& 4.54 & 2.14 & 3.65 & 10.33
		\\
		S11 &  &  & 32s / 2s
		& 3.51 & 2.23 & 2.72 & 8.46
		& 10.15 & 1.56 & 8.43 & 20.14
		& 4.44 & 2.14 & 3.52 & \textbf{10.10}
		\\
		S12 &  &  & 64s / 2s
		& 3.47 & 2.44 & 2.74 & 8.65
		& 9.76 & 1.76 & 8.38 & 19.90 
		& 4.35 & 2.35 & 3.53 & 10.23
		\\
		\bottomrule
	\end{tabular}
\end{table*}

When only loading audio-related modules (green-colored in Fig.~\ref{fig_mimo}), the inference \textit{stage 1} of MIMO-TSVAD is equivalent to the basic Seq2Seq-TSVAD~\cite{cheng2023target}. Using the pre-trained speaker embedding model, we implement the AHC~\cite{wang2022dku}, LSTM-SC~\cite{wang2022similarity}, and VBx~\cite{landini2022bayesian} as modularized diarization methods to extract initial speaker profiles. The related hyper-parameters are tuned on the development sets of VoxConverse~\cite{chung2020spot} and DIHARD-III~\cite{ryant2020third} datasets, respectively.
		
For the VoxConverse dataset, the AHC obtains a Diarization Error Rate (DER) of 5.35\% on the test set with a tolerance collar of 250 ms, better than the VBx (5.62\%) and LSTM-SC (6.33\%). Thus, Table~\ref{exp:voxconverse} illustrates the performance of different MIMO-TSVAD models initialized by the AHC result. Two types of VAD resolutions (duration per frame-level prediction) are provided as coarse (80 ms) and precise (10 ms) options, which can be implemented easily by adjusting the dimension of the model output. As the dataset has up to 21 speakers in a single recording, we choose the speaker capacity (maximum speaker embeddings) of 10, 20, and 30 to cover insufficient, suitable, and sufficient capacities, respectively. Systems S1-3 reveal that total DERs reduce obviously from 7.36\% to 4.98\% when the speaker capacity increases from 10 to 30. The gain mainly comes from the subset of recordings with over 10 speakers. Systems S4-6 show that the precise resolution (10 ms) slightly improves over a coarse resolution (80 ms), except System S6. It is speculated that the tolerance collar makes the evaluation insensitive to utterance boundaries. Based on the VAD resolution of 10 ms and speaker capacity of 30, Systems S7-9 increase the chunk length of model training to decrease total DERs from 4.90\% to 4.53\%. Systems S10-12 utilize a 2-second chunk shift for inference. Beyond simply stitching chunk-wise predictions, overlapped regions are averaged as score-level fusion. It can be seen that dense inference can further decrease the lowest total DER to 4.18\%. Table~\ref{voxconverse_comp} compares our proposed method with the current state-of-the-art results. Our best performance significantly outperforms previous ones, especially for some multi-system fusion results in early VoxSRC challenges. Notably, System S12 participates in the latest VoxSRC-23 winning system~\cite{cheng2023dku}. Although the pyannote.audio~\cite{baroudi2023pyannote} team achieves a DER of 4.00\% based on large-scale data and unsupervised WavLM models, it does not beat our winning system on the VoxSRC-23 challenge set.

For the DIHARD-III dataset, reference VAD is provided according to the challenge's track 1 rules. The LSTM-SC obtains the 15.40\% DER on the evaluation set, better than the VBx (16.58\%) and AHC (16.77\%). Thus, Table~\ref{exp:dihard} illustrates the performance of different MIMO-TSVAD models initialized by the LSTM-SC result. Since the number of speakers in each recording is up to 9, we explore the speaker capacity of 5, 10, and 20. The experimental results show similar conclusions to Table~\ref{exp:voxconverse}, which means the precise VAD resolution, larger speaker capacity, longer chunk length, and dense inference can usually enhance the diarization performance. Nevertheless, two phenomena should be mentioned. First, the benefits of precise VAD resolution are more significant than the maximum speaker capacity here. As the DIHARD-III dataset is annotated at 10 ms and evaluated without the tolerance collar, its results are more likely affected by the temporal precision of estimated speech activities. Second, the diarization performance saturates when the chunk length grows to 32 seconds in System S8 and S11, which reveals that the performance cannot be further improved easily. Finally, System 11 achieves the lowest total DER of 10.10\%. Table~\ref{dihard_comp} compares our proposed method with the current state-of-the-art results. Our best performance demonstrates superiority over existing approaches.

\begin{table}[t]
	\centering
	\setlength{\tabcolsep}{18pt}
	\caption{Comparisons of MIMO-TSVAD models with others on the DIHARD-III evaluation set (Reference VAD).}
	\label{dihard_comp}
	\begin{threeparttable}[b]
	\begin{tabular}{lr}
		\toprule
		\textbf{Method} & \textbf{DER (\%)} \\
		\midrule
		Hitachi-JHU~\cite{horiguchi2021hitachi}~\tnote{$\dag$} & 11.58  \\
		USTC-NELSLIP~\cite{wang2021ustc}~\tnote{$\ddag$}  & 11.30  \\
		Wang et al.~\cite{wang2021scenario} & 11.30\\
		ANSD-MA-MSE~\cite{he2023ansd} & 11.12 \\
		\midrule
		AHC~\cite{wang2022dku} &  16.77\\
		VBx~\cite{landini2022bayesian} &  16.58 \\
		LSTM-SC~\cite{wang2022similarity}  & 15.40 \\
		\quad $+$ MIMO-TSVAD (S11 in Table~\ref{exp:dihard}) & \textbf{10.10} \\
		\bottomrule
	\end{tabular}
	\begin{tablenotes}
		\item[$\dag$] DIHARD-III 2nd-ranked result with 5-system fusion.
    	\item[$\ddag$]  DIHARD-III 1st-ranked result with 5-system fusion.
    \end{tablenotes}
	\end{threeparttable}
\end{table}

Notably, the DER results of MIMO-TSVAD models in Tables~\ref{exp:voxconverse} and~\ref{exp:dihard} differ slightly from those reported in the original Seq2Seq-TSVAD paper~\cite{cheng2023target}. A clarification of the inference configurations is provided as follows.
\begin{itemize}
  \item First, for the VoxConverse dataset, this work does not apply any post-processing based on estimated VAD information, whereas \cite{cheng2023target} revises TS-VAD predictions using results from a separate diarization system. In this case, we observed that a finer-grained inference strategy, achieved by adjusting chunk lengths and chunk shifts, already leads to significant DER improvements. As a result, the best-performing MIMO-TSVAD model achieves competitive accuracy without requiring additional VAD-based refinement.
  \item Second, for the DIHARD-III dataset, \cite{cheng2023target} directly reports the best DER under a specific configuration (chunk length = 16 s, VAD resolution = 10 ms, speaker capacity = 20). In contrast, this work systematically explores a wide range of inference setups to gain a deeper understanding of their impact on performance.
  \item Third, the reported DERs in this work are averaged over three independent runs to mitigate randomness, whereas \cite{cheng2023target} reports single-run outcomes. These differences may account for the observed variations in DER.
\end{itemize}

\subsection{Evaluation of Video-based and Audio-Visual Diarization}

\begin{table*}[t]
	\centering
	\setlength{\tabcolsep}{5pt}
	\caption{Performance of different MIMO-TSVAD models on the MISP 2022 evaluation set (Reference VAD). The Diarization Error Rate (DER) is the sum of miss (MI), false alarm (FA), and speaker confusion (SC) rates.}
	\label{exp:misp}
	\begin{threeparttable}[b]
	\begin{tabular}{llllllllrrrr}
		\toprule
		\multirow{2}{*}{\textbf{ID}} 
		&\multirow{2}{*}{\textbf{Inference}}
		&\multicolumn{3}{c}{\textbf{Input}}
		&\multicolumn{3}{c}{\textbf{Output}}
		&\multirow{2}{*}{\textbf{MI (\%)}}
		&\multirow{2}{*}{\textbf{FA (\%)}}
		&\multirow{2}{*}{\textbf{SC (\%)}}
		&\multirow{2}{*}{\textbf{DER (\%)}}	
		\\
		\cmidrule(lr){3-5} \cmidrule(lr){6-8}
		& 
		& \textbf{Audio}
		& \textbf{Video}
		& \textbf{Audio-visual}
		& \textbf{Audio-based}
		& \textbf{Video-based}
		& \textbf{Mixed}
		& 
		\\
		
		\midrule
		S1 & Stage 1  & \checkmark & & & \checkmark & & 
		& 7.85 & 2.01 & 13.49 & 23.35	
		\\
		S2 & Stage 2  &   & \checkmark &  &  &\checkmark &
		& 7.36 & 3.31 & 4.34 & 15.01
		\\
		S3 & Stage 3  &  & & \checkmark &  &  \checkmark & 
		& 4.77 & 2.10 & 3.19 &  \textbf{10.06}
		\\
		\midrule
		
		\multirow{5}{*}{S4~\tnote{$\ast$}}
		& Stage 1  & \checkmark & & & \checkmark & & 
		& 8.25 & 2.85 & 14.75 & 25.85
		\\
		& Stage 2  &   & \checkmark &  &  &\checkmark &
		& 9.89 & 1.53 & 5.16 & 16.58
		\\
		& Stage 3  &  & & \checkmark &  &  \checkmark  & 
		& 4.60 & 2.48 & 3.31 &  \textbf{10.39}
		\\
		\cmidrule(lr){2-12}
		& Stage 4 w/o SA~\tnote{$\bigtriangledown$}  & & &  \checkmark & & &  \checkmark 
		& 3.84 & 2.69 & 2.25 & 8.78
		\\
		& Stage 4 w/ SA~\tnote{$\bigtriangledown$} & & &  \checkmark & & &  \checkmark  
		& 3.84 & 2.68 & 2.22 & \textbf{8.74}
		\\
		
		\midrule

		\multirow{5}{*}{S5}
		& Stage 1  & \checkmark & & & \checkmark & & 
		& 9.05 & 1.98 & 12.95 & 23.98
		\\
		& Stage 2  &   & \checkmark &  &  &\checkmark &
		& 7.45 & 2.74 & 4.22 & 14.41
		\\
		& Stage 3  &  & & \checkmark &  &  \checkmark  & 
		& 4.65 & 2.43 & 3.08 & \textbf{10.16}
		\\
		\cmidrule(lr){2-12}
		& Stage 4 w/o SA~\tnote{$\bigtriangledown$}  & & &  \checkmark & & &  \checkmark 
		& 3.97 & 2.55 & 1.68 & 8.20
		\\
		& Stage 4 w/ SA~\tnote{$\bigtriangledown$} & & &  \checkmark & & &  \checkmark  
		& 3.93 & 2.56 & 1.66 & \textbf{8.15}
		\\
		\bottomrule
	\end{tabular}
	\begin{tablenotes}
    	\item[$\ast$] indicates the model trained without the proposed multi-stage training strategy.  
    	\item[$\bigtriangledown$] denotes the abbreviation of speaker alignment.
    \end{tablenotes}
	\end{threeparttable}
\end{table*}

When loading all modules in Fig.~\ref{fig_mimo}, the MIMO-TSVAD can conduct various inference stages flexibly. Limited by the large GPU memory consumption of multi-modal data, we only train the models under the VAD resolution of 10 ms, speaker capacity of 6, chunk length of 8 seconds, and chunk shift of 2 seconds. Also, we utilize the first channel of far-field audio in the MISP 2022 training set to tune the AHC~\cite{wang2022dku}, LSTM-SC~\cite{wang2022similarity}, and VBx~\cite{landini2022bayesian}, respectively.

For the MISP 2022 evaluation set, reference VAD is provided according to the challenge's track 1 rules. The LSTM-SC obtains the 29.30\% DER, better than the AHC (30.02\%) and VBx (33.37\%). Thus, Table~\ref{exp:misp} illustrates the performance of different MIMO-TSVAD models initialized by the LSTM-SC result. To investigate the potential of audio-based, video-based, and mixed output methods, Systems S1-3 are individually trained for each inference stage. Experimental results show that the video-based System 2 obtains 15.01\% DER, surpassing the audio-based System 1 with 23.35\% DER. The improvement mostly comes from the speaker confusion (SC) error decreasing from 13.49\% to 4.34\%. Visual modality demonstrates a significant advantage in determining speaker identities as it does not suffer from the issue of overlapping speakers. Then, System 3 decreases the DER to 10.06\% by introducing auxiliary audio information. About $3/4$ improvement (3.8\% of 4.95\% DER) is contributed by miss (MI) and false alarm (FA) rates. Compared with video signals of a limited 25 FPS, it can be seen that the audio modality can bring higher temporal precision for predicted utterance boundaries.

\begin{table}[t]
	\centering
	\setlength{\tabcolsep}{18pt}
	\caption{Comparisons of MIMO-TSVAD models with others on the MISP 2022 evaluation set (Reference VAD).}
	\label{misp_comp}
	\begin{threeparttable}[b]
	\begin{tabular}{lr}
		\toprule
		\textbf{Method} & \textbf{DER (\%)} \\
		\midrule
		 E2E-AVSD (Official Baseline)~\cite{he2022end}  & 13.88 \\
		NPU-FlySpeech~\cite{zhang2023flyspeech}~\tnote{$a$} & 10.90  \\
		SJTU~\cite{liu2023multi}~\tnote{$b$} & 10.82 \\
		WHU-Alibaba~\cite{cheng2023whu}~\tnote{$c$} & 8.82 \\
		\midrule
		VBx~\cite{landini2022bayesian} & 33.37 \\
		AHC~\cite{wang2022dku} & 30.02 \\
		LSTM-SC~\cite{wang2022similarity} & 29.30  \\
		\quad $+$ MIMO-TSVAD (S5 in Table~\ref{exp:misp}) & \textbf{8.15} \\
		\bottomrule
	\end{tabular}
	\begin{tablenotes}
    	\item[$a,b,c$] denote the 3rd-, 2nd-, and 1st-ranked submissions to the audio-visual diarization track of MISP 2022 Challenge, respectively. 
    \end{tablenotes}
	\end{threeparttable}
\end{table}

However, the independent training of Systems S1-3 is not cost-effective. Accordingly, we propose a multi-stage training strategy to obtain an integrated model for all inference stages. Once trained, the model supports all the functionalities of Systems S1-3 and unlocks the new mixed output method. We first train System S4 starting from the last stage described in Section~\ref{sec:training}. Without the multi-stage design, experimental results show that although the inference \textit{stages 1-3} of System 4 degrade considerably compared with the counterpart Systems 1-3, the inference \textit{stage 4} still reduces the DER to 8.74\% by adopting the advanced mixed output method. Then, we train System S5 with the presented multi-stage training strategy as shown in Fig.~\ref{fig_training}. It can be seen that the performance degradation of inference \textit{stages 1-3} is alleviated obviously. Meanwhile, its inference \textit{stage 4} reaches the lowest DER of 8.15\%. As no dramatic modality-missing problem exists in the MISP 2022 dataset, the benefits of adopting the Speaker Alignment (SA) module are not noteworthy here but are thoroughly investigated in the next section. Furthermore, Table~\ref{misp_comp} compares our proposed method with the current state-of-the-art results. The MIMO-TSVAD method updates our previous system~\cite{cheng2023whu} that has won the audio-visual diarization track of the MISP 2022 Challenge.

\subsection{Evaluatioin of Robustness to Lip-Missing Scenarios}

A speaker's lip track may not always be available in real scenarios. To explore the impact of lip-missing problems on the MIMO-TSVAD framework, we newly simulate test data based on the original MISP 2022 evaluation set. Three lip-missing scenarios are created using zeros to randomly mask the Lip-RoI for simulating off-screen data, described as follows.
\begin{itemize}
  \item \textit{Partially Off-screen}: Each speaker's Lip-RoIs are removed during a period. The total number of speakers in the video remains unchanged.
  \item \textit{Completely Off-screen}: The Lip-RoIs of some selected speakers are entirely removed. Fewer speakers exist in the video. 
  \item \textit{Hybrid}: Both situations above may happen.
  \end{itemize}

\begin{figure*}[t]
\centering
	\subfloat[Partially Off-screen]{\includegraphics[width=2.2in]{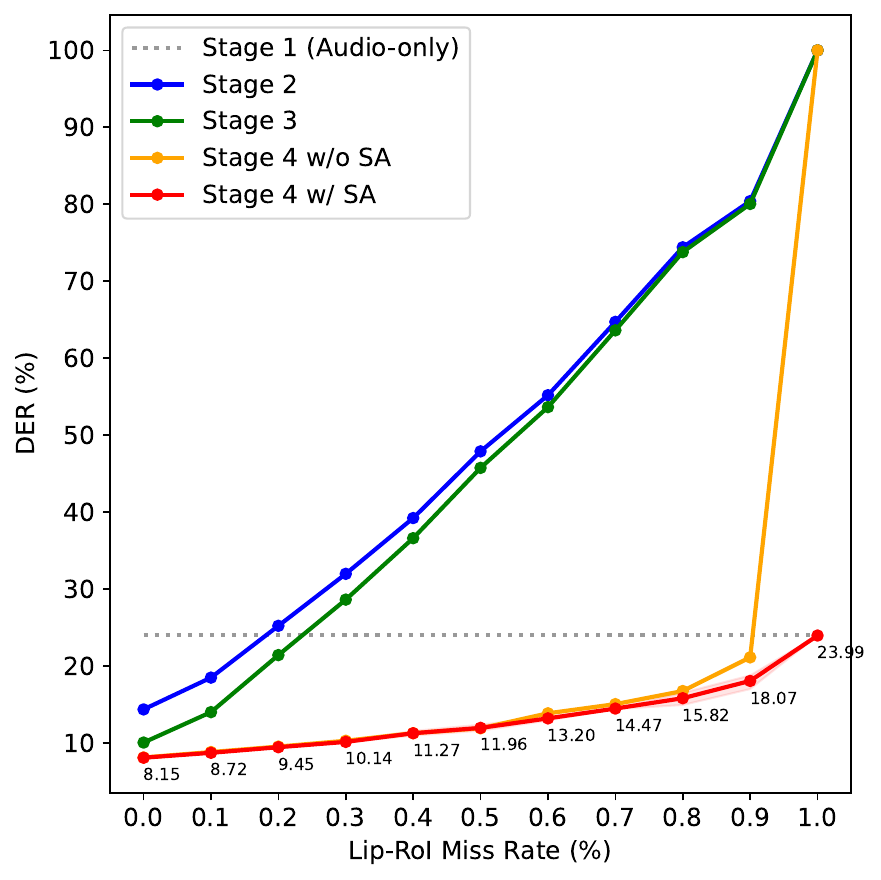}
	\label{fig_miss_p}}
	\hfil
	\subfloat[Completely Off-screen]{\includegraphics[width=2.2in]{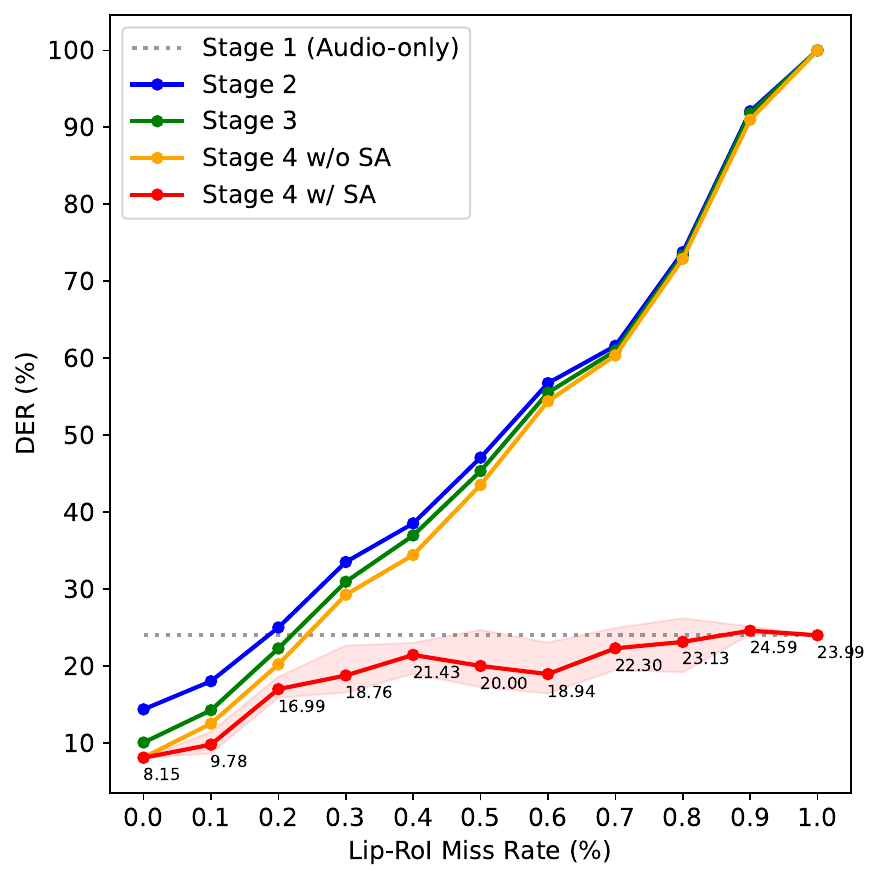}
	\label{fig_miss_f}}
	\hfil
	\subfloat[Hybrid]{\includegraphics[width=2.2in]{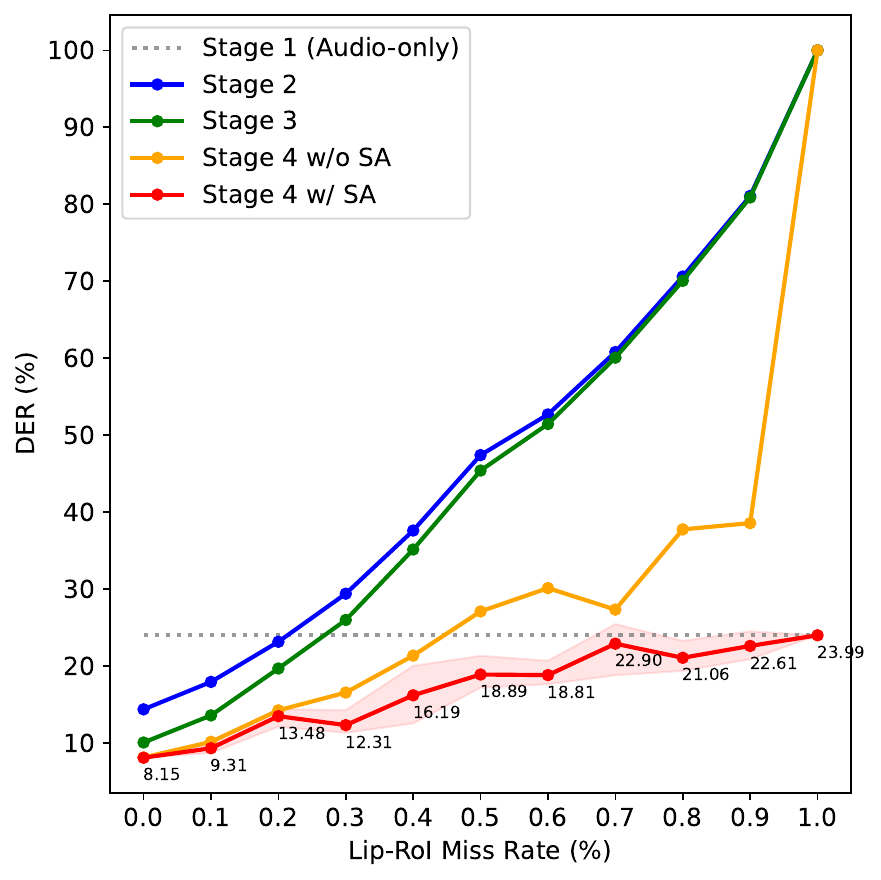}
	\label{fig_miss_h}}
\caption{DERs (\%) of MIMO-TSVAD models on the simulated MISP 2022 evaluation sets with different Lip-RoI miss rates. Each value is the average performance on three copies of the independent simulation, where the red-shaded bands indicate the maximum-minimum intervals.}
\label{fig_robust}
\end{figure*}

Fig.~\ref{fig_robust} demonstrates how the performance of the MIMO-TSVAD system (S5 in Table~\ref{exp:misp}) changes under different lip-missing scenarios and degrees. Since the inference \textit{stages 2-3} entirely rely on the video-based output, their DERs increase almost linearly as the Lip-RoI miss rates rise in all scenarios. Notably, the inference \textit{stage 4} without speaker alignment performs differently in different lip-missing scenarios. In Fig.~\ref{fig_miss_p}, it utilizes the mixed output method to keep considerable robustness as long as each speaker's few Lip-RoI segments can successfully extract the paired speaker embedding. Once the Lip-RoI miss rate reaches 100\%, the system directly fails. In Fig.~\ref{fig_miss_f}, its DERs also increase rapidly with the Lip-RoI miss rate increases because all the missed Lip-RoI comes from completely off-screen speakers whose speaker embeddings cannot be enrolled by the video-based output of the previous stage. In Fig.~\ref{fig_miss_h}, its DER curve combines the characteristics in Fig.~\ref{fig_miss_p} and Fig.~\ref{fig_miss_f}. The inference \textit{stage 4} without speaker alignment may recall partially off-screen speakers' voice activities but is useless for completely off-screen speakers. Lastly, using the speaker alignment module dramatically improves the robustness of inference \textit{stage 4}. With the help of undetected speaker embeddings, the mixed output method can work stably in different cases.

In general, our proposed MIMO-TSVAD framework exhibits strong robustness to complex lip-missing scenarios. With the Lip-RoI missing rate varying from 0 to 100\%, it transits from an audio-visual to an audio-only system, maintaining the DER within a satisfactory range.

\subsection{Computing Efficiency}
Table~\ref{efficiency} illustrates the computing efficiency of MIMO-TSVAD models trained in Table~\ref{exp:misp}. For each 8-second input data, several metrics for the different inference stages are presented as follows. First, the number of parameters for each stage is an essential indicator. Second, Floating-Point Operations (FLOPs) are used to measure the computational complexity. Then, we count the required GPU memory and the average time for inferring each input chunk, which is tested on the NVIDIA RTX-3090 GPU. Starting from the second stage, the use of visual modality significantly increases the computational load. The more multi-modal information the model uses, the greater the computational load is required. Improving the computing efficiency of multi-modal speaker diarization systems is still challenging.

\begin{table}[t]
	\centering
	\setlength{\tabcolsep}{5pt}
	\caption{Computing Efficiency of MIMO-TSVAD models for each 8-second input
data in different inference stages, regarding the number of parameters, Floating-Point Operations (FLOPs), GPU memory, and inference time.}
	\label{efficiency}
	\begin{threeparttable}[b]
	\begin{tabular}{lrrrrr}
		\toprule
		\textbf{Stage} & \textbf{Params (M)} & \textbf{FLOPs (G)} & \textbf{Memory (MB)} & \textbf{Time (s)}\\
        \midrule
        1 & 76.56  & 151.80 & 494.49  & 0.0279 \\
        2 & 64.11  & 515.60 & 943.60  & 0.0499 \\
        3 & 85.78  & 667.01 & 1040.50 & 0.0617 \\
        4 & 153.72 & 669.16 & 1313.22 & 0.0797 \\
		\bottomrule
	\end{tabular}
	\end{threeparttable}
\end{table}

\section{Conclusions}
\label{sec:conclusions}

This paper proposes a novel MIMO-TSVAD framework to tackle speaker diarization under complicated audio-visual data accessibilities. The model with jointly designed multi-stage training and inference strategies is compatible with different scenarios in a unified framework. Experimental results show that the MIMO-TSVAD framework performs well for audio-based, video-based, and audio-visual speaker diarization. It obtains new state-of-the-art DERs of 4.18\% on the VoxConverse~\cite{chung2020spot} test set, 10.10\% on the DIHARD-III~\cite{ryant2020third} evaluation set, and 8.15\% on the MISP 2022~\cite{wang2023multimodal} evaluation set, respectively. Furthermore, the MIMO-TSVAD framework demonstrates strong robustness against lip-missing problems. In simulated scenarios with varying lip-missing degrees, it guarantees that the DERs of the audio-visual system are always no worse than the audio-only system. In the future, we will further improve the current approach regarding advanced lip extraction and clustering methods, better use of multi-channel audio, etc.

\section*{Acknowledgments}
This research is funded in part by the National Natural Science Foundation of China (62171207), Yangtze River Delta Science and Technology Innovation Community Joint Research Project (2024CSJGG01100), Science and Technology Program of Suzhou City (SYC2022051) and Guangdong Science and Technology Plan (2023A1111120012). Many thanks for the computational resource provided by the Advanced Computing East China Sub-Center.

%
\bibliographystyle{IEEEtran}
\bibliography{refs}


\end{document}